# Application of Optimization and Simulation to Musical Composition that Emerges Dynamically during Ensemble Singing Performance

**Alexis Kirke** (corresponding author), Interdisciplinary Centre for Computer Music Research, University of Plymouth, Drake Circus, Plymouth, PL4 8AA, UK, alexis.kirke@plymouth.ac.uk

**Greg B. Davies**, Head of Behavioural Finance, Oxford Risk Research & Analysis Ltd, 1 Paris Garden, London SE1 8ND, UK

**Joel Eaton**, Senior Instructional Designer, ARM, Seaford, East Sussex, UK.

This paper presents and tests a new approach to composing for ensemble singing performance: "reality opera". In the performance of such a composition, emotions of the singers are real and emerge as a consequence of their interactions and reaction and to a dynamic narrative. This paper gives background and motivation for the form, based on three key concepts, incorporating the use of technology. Then proposed techniques for creating reality opera are instantiated in an example, which is performed and a behavioural analysis done of performer reactions, leading to support for the feasibility of the reality opera concept.

## Introduction

Opera is most simply defined as: "an art form in which singers and musicians perform a dramatic work combining text (called a libretto) and musical score, usually in a theatrical setting." Such an art form would seem incompatible with the concept of reality TV: "a genre of television programming that documents unscripted situations and actual occurrences". There are two seemingly insurmountable differences for the creator, aside from the setting. The first is that the opera is pre-scripted, whereas the reality show is only partially scripted. The second difference is that the emotional trajectory of a performer in an opera is acted and is not real – they take on a character based on the libretto and score. Whereas the emotions show in reality television are real emotions, defined by largely unpredictable interactions. Of course, reality television creators do set up scenarios and constraints to manipulate emotional trajectories, and use editing and shot choice to emphasize the emotions they want (Turner, 2006). Providing singers with greater autonomy in opera performance through technology has been seen as a way of challenging traditional power structures (Unander-Scharin, 2015). Technology has also been used in opera to increase audience engagement with the characters and narrative (Machover, 2011). However such technology-led innovations in opera can be controversial (Henson, 2016) and so should be examined through testing with audience and performers. In this paper, the application of technology to enable a new genre of opera is proposed, based on the real emotions and emerging dynamics amongst a cast playing themselves. Three key elements are presented as the basis of this form, called "reality opera". The opera form is tested through an instantiation in an example performance from which data was collected.

Tightly constrained musical non-deteministic vocal performances have many predecessors as shown in (Nyman, 1974) . Outside of opera, improvisational drama obviously has a long history (Chilver, 1967). This drama can also take on a "reality" element where the actors enter into real relationships with each other or the audience (Rand, 1935)(Econn, 2007). Within opera there is some history of improvisation with Cantonese opera (Yun, 1989), and obviously with Cadenzas (Siegwart et al., 1995) and da capo arias, as well as some technology-enabled interactive opera (Bonardi et al., 2002). Key innovators in the field of opera have been Machover's group at MIT (Earley, 2014). For example the Brain Opera (Machover, 1996) is an installation in which a large number of interactive experiences - such as singing trees and a gesture wall - interact with the audience to create an overall musical experience.

Even in traditional opera, performers do adjust their performances to interpret the emotions of the character, and will become immersed in the character to a degree, but sill stay true to a fixed score. However these do not come close to being a singing drama which impacts on the real feelings of the performers, feeding back into significant changes in the emerging drama and music. It would seem that to create something meaningful, constraints need to be added, like fixed music or fixed libretto. But then the freedom of the performers to develop a genuine emotional trajectory of their own is lost.

In "reality opera", through a selection of story-type and constraints, it is possible to incorporate a number of elements of the musical and dramatic, while providing some of the underlying undetermined nature of reality television. We propose a number of solutions to crossing the barrier between opera and reality opera: (i) the restriction of the narrative to situations involving formalized human interactions, (ii) the use of methodologies from machine learning to develop and experiment with the constraints, and (iii) the use of computer simulation to test the results during composition and to incorporate dynamic elements of the outside world into the narrative.

Looking first at item (i), there are a number of situations that formalize human interactions but which could be of narrative interest. Examples include stock trading, speed dating, and various sports or games. Consider speed dating. In this males and females are required to interact for a certain period of time, and tend to use similar questions. It is natural to make this musical, either by considering the use of song in sexual relationships (Dissanayake, 2008), or by considering the metaphor of bird song – in which birds sing to attract mates as well as to mark territory (Earp et al., 2012). This "making it musical" is a required creative step in the process of utilizing the reality opera concept. There must be a creative leap that links the normal human situation with a sung situation: providing the audience with a rationale.

Item (ii) above – the use of constraints – relates to what exactly the daters might sing. To enable them to become genuinely caught up in the process, they need to have agency – the ability to choose what they sing and to whom. Implicit in this is a major constraint. If all singers are able to choose what and when they sing – as required in reality opera - the resulting musical sound could be incredibly unpleasant. Thus various constraints would be required of the composer: for example, that the various sung lines need all overlap and harmonise with each other in a controlled way. For a large number of sung lines, a human composer may lost the ability to ensure co-harmonisation. This points towards the possible use of a technology such as harmonic optimization software.

The second part of Item (iii) – the testing of results during composition – could involve the composer manually generating a series of scores with different combinations of the sung lines, and listening to each. Once again, for a large number of lines and a – say – 60 minute opera, this could be an unfeasible process. Thus pointing once again to the potential usefulness of computer music techniques.

The second part of Item (iii) – the external world - is less relevant in the speed dating reality opera, as the external world does not interfere so much with a speed dating session in a bar. Most circumstances are implicit in the past experience and desires of the performers who are dating. One requirement / constraint this might raise is that all performers be single in the real world. This would lead to them being open to the possibility of true engagement in the meaning of the performance.

An element to consider when discussing the concept of reality opera, which is not covered in (i) to (iii), is the cognitive load on performers. If performers are working within a set of constraints to fulfill (i) to (iii) and also have the implicit constraint of performing well, they will have a significant cognitive load. There had been research into the effects of cognitive load on musical performers (Corlu et al., 2015a), including work specific work in opera (Corlu et al., 2015b). Thus any investigation of reality opera needs to consider this issue, as cognitive load might degrade performance or cause the performers to not become truly immersed in their emotions.

The investigation will be initiated through investigation of a specific instantiation of a reality opera. This provides a way of testing the concept: i.e. can the performers truly become emotionally involved in the performance, based on the narrative they help to define? We will present the example reality opera set in a stock market. (A key reason for choosing the stock market instantiation was that the first author worked as a Quantitative Analyst for a Wall Street broker before becoming a composer, thus placing it within an area of past expertise.) The example will show explicitly applications of techniques inspired by the above items (i) to (iii), and the data resulting from it. Before doing so, we will examine other related musical work in the stock market world.

**Related Work**

"Open Outcry" is set in the world of stock markets and market data. Past musical work involving such markets and data does exist. Various sonic representations have been developed either for creative reasons, or for sonification (Cohen, 1994) to allow audio display of data for traders and analysts. Examples include the Dow Piano (Edland, 2010), which represents the trading in 2010 by converting trading volumes to loudness and pitch to daily closing prices. Worrall (2009) uses sounds to try to identify correlations in stock markets. And (Ciardi et al., 2013) uses sonification to provide eyes free display for traders. A slightly different approach is utilized in (Kirke et al. 2012), mapping the market returns and trading volumes onto a market "emotional state", and then sonifying this emotional state with music which communicates the relevant emotion. For example: high tempo major key music for a "happy" market.

There have also been examples of sonic representations from the financial markets for entertainment. For example Ear-trading (Moon et al. 2010) is a game for the iPhone which uses psychoacoustic surround sound to sonify market data can help users to trade in a game more successfully. Pitches going down mean the price is going down, upward means price is going up. Tempo maps to the trading volumes - higher tempo means more trading is happening. Different stocks are represented by different timbres. The surround sound is designed to allow for more streams of data to be audible. In "Notes on a Crisis" (Studemann, 2010) composer Julian Anderson provided a journalist with initial notes for how he would use musical structure to express the financial crisis of 2008-2009. For example just before the crash "the sound of a high-pitched 'minimalist machine', creating a galloping rhythm under which bass notes hint at trouble ahead." Then as the crash starts "the top-of-the-market high note is moved up a notch – then held. Alongside it he launches a series of step-by-step descents using the original chord, which is gradually transposed downward."

Other recreational examples include Playing the Market (Emerald Suspension, 2006) - an album by Emerald Suspension which includes tracks composed based on patterns in the stock markets and in economic data. For example "Long Bond" is a string piece based on the pattern of interest rates since 1926 as represented by the yield of the 30-year Treasury bond. High interest rates map onto high pitches, and low interest rates onto low pitches. Outside of music, the stock market has been the inspiration for dynamic performances. The play "Dead Cat Bounce" allowed the audience to select stocks which were traded live during the play, and which they benefited from (Kondek,

2005). However there have been no dynamic musical performances based around singers trading stocks.

A final area that relates to the work presented here, is musical games. These have a long history (Harr, 1962). More recent examples are Xenakis' musical games of Duel and Strategie (Luini et al., 2006). These involve conductors of two orchestras playing against each other. A conductor makes a move by choosing a part of the score and getting his orchestra to play it to the second conductor. The second conductor then responds with a part of his score. The second conductor needs to select a part of the score which maximizes their chance of winning. This chance is calculated by the rules of the particular game defined by Xenakis.

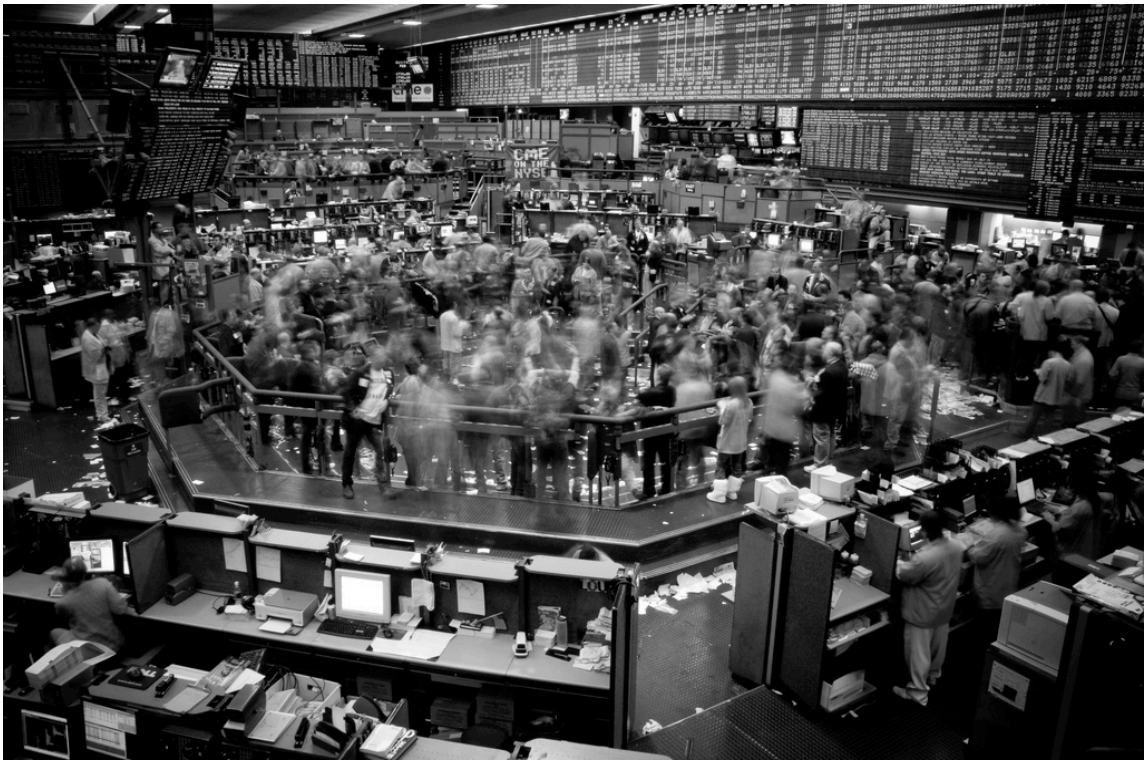

**Figure 1.** An Open Outcry Trading Floor.

**Instantiation of Reality Opera**

A suitable scenario for reality opera is a structured human interaction in which the performers can truly participate emotionally and musically, and one that can be mediated by semi-constrained musical interaction. The scenario chosen for the example in this paper is a particular type of financial market called an open outcry market, as example of which is shown in Figure 1. There are still major international financial markets in which trading is done by shouting and waving – known as open outcry trading. To gain some understanding of the market state, one does not need to employ advanced statistical analysis algorithms. Simply stand in the room like a member of an audience watching a show. Look at the faces of the traders; listen to the sounds of their voices. Is the trading floor roaring or silent? Do you sense fear or jubilation? In the same way that singing works as a creative metaphor for speed dating interaction, for reasons described earlier, the expansive vocal expression on an open outcry trading floor is emotional and performative. The trader is performing for other traders to get a reaction, there are rhythms of call and response, and the trader's emotions are hard to hide in more extreme market situations – it is a highly expressive environment. All this provides a clear and convincing and metaphorical transformation into emotional and communicative ensemble singing, and thus is another element of its suitability for reality opera.

The story is about 12 open outcry traders. They have been provided with a portfolio of fictional stocks, and some fictional money. They are able to freely trade these stocks amongst themselves by singing – without any constraints as to when to sing - from a libretto which allows them to define what trades happen when. Although the stocks and money are fictional, the value is not. At the end of the opera the singer's fee will include a bonus based on how much money they made in their trading. Thus in addition to their emotional immersion created simply by the natural competitive desire, they are motivated by the actual desire for money.

The question then becomes how to embed these choices and behaviour in music, as well as to impose a dramatic arc of some sort. There is a dramatic arc implicit in the financial markets – they are cyclical, with periods of boom and bust. In this reality opera, the market in which the performers trade is controllable by the conductor in a way not visible to the performers. The conductor can make the underlying market model crash, boom or remain steady. Different states have a different impact on the performers.

**Music**

Looking to studies in music and emotion it can been seen that for western listeners, minor keys tend to communicate lower emotional positivity, and major keys higher emotional positivity (Juslin et al., 2009). Similarly in terms of harmony it is found that consonant harmonies tend to express more positive emotions than dissonant harmonies (Cook et al., 2008). Based on this, a strategy was developed.

For each of the fictional stocks, there is be a "buy tune" and a "sell tune" that the singer has to memorize. They can insert different quantities and prices into the wording, thus altering the rhythms and emphases slightly, but the tune pitch-classes remain the same. The buy tunes are composed so that they harmonise each with the other in a consonant way. The sell tunes are written so that they are relatively more dissonant when sung together. Finally the buy and sell tunes for the same simulated stock are written so as to harmonize pleasantly. This is designed so that if most people in the market are trying to buy stocks, there will be consonance, whereas if most people in the market were trying to sell there is dissonance. Thus the selling market – the downward-moving market – is more dissonant than the buying market – the upward-moving market. The harmonies are designed to communicate the way that people normally feel about markets. A bull market is considered positive, a crashing market negative.

The performers can sing the tunes at any point of overlap with each other. This means that one tune could be started halfway through another tune. The process of creating the structure of dissonance and consonance could be complex. This can be addressed using computer music optimization techniques, for example – as done here - genetic algorithms.

**Trading Tunes**

Tunes were composed to implement the three harmonic constraints already described for the trading phrases. Genetic algorithms (GA) (Goldberg, 1989) have been used many times in algorithmic composition (Ting et al., 2017)(Waschka, 2007) and were used to help in composing the phrases. Only an overview will be given here, as the GAs turned out to not be so necessary in this example "reality opera". The use of GAs was initially due to the expectation that there would be 8 or more stocks, which would require at least 16 trading phrases, all different, and harmonizing in specific ways. However, during discussions and workshopping with singers it became clear that the number of phrases needed to be limited for cognitive load purposes, leading to a total of 3 stocks. Thus the tunes could have been written manually, but GAs were still used, as they were seen as useful because it enabled the rapid creation of multiple candidate sets of trading tunes which could be tested.

Tunes of four notes (the length is explained later) were created initially. They had random pitches and random quantized timings. Buy tunes were initialized in C major, sell tunes in C minor. Each GA population member consisted of 3 buy tunes and 3 sell tunes – because 3 artificial stocks are used in the opera. A population of 120 was used in the GA and iterated through 4000 cycles. The fitness function used was:

$$Fitness = 1.5*buySellConsonance + buyBuyConsonance$$

$$- sellSellConsonance \qquad (1)$$

For artistic reasons the consonance between buy and sell phrases was most highly weighted in the calculation, by multiplying by 1.5. Consonance is a measure of how well phrases harmonize together. A low consonance means a higher musical dissonance. The mathematical function used to measure consonance was developed based on the formal rules of harmony (Jamini, 2005) and then freely adapted to the composer's ear. For example, the most dissonant measurements were given to the tritone, semitone intervals and major 7th. Some key differences from the formal rules of harmony included that major thirds were scored less dissonant than fifths. Consonance scores of 6, 5, 3 and 1 could be awarded to intervals, higher values being more consonant. To find the consonance for a set of tunes (e.g. buy tunes sung against sell tunes) the consonance of all buy tunes for a population member against all sell tune for that member were calculated pair-wise and averaged. Furthermore, the "shift-scores" were incorporated into the average. These involved shifting one tune in time by 1, 2, or 3 notes and calculating the consonance. The reason for this was that in opera interaction between multiple singers, harmonized trading tunes would not necessarily be in phase.

      At each GA generation the top 20% scoring members were allowed to breed. Breeding pairs were picked randomly from these, as was the gene split point for combining. The genes were simply linear musical note lists, so the first part of one tune would be added to the end part of another tune to breed a new population member. During this process there was a 25% chance of mutation. Mutation was a fairly brutal

affair. The entire buy and sell pair for a particular stock in a particular population member would be replaced with new randomly generated buy and sell tunes. After 4000 iterations the highest scoring population member was selected.

The testing was as follows. Six candidate sets of trading tunes were generated by the GA, whose constraints implemented the ideas already discussed. A musical multi-agent system (Kirke et al., 2015) was created to simulate the sound world of the opera based on a given trading tune set. It consists of agents that at any time step may randomly decide to buy or sell one of the three stocks. They then repeat the trading tune a number of times. The trading tune would be adjusted to include different numbers of syllables, since different prices and amounts would require more or fewer notes to be used. The simulation was designed so that at times more agents would be buying than selling and more agents selling than buying – so as to simulate booms and crashes. Also there was a density variable which allowed trading activity to be increased and decreased over the period of the simulation. The resulting music of 12 agents in a simulation was output as a MIDI file which was imported into a singing synthesizer. For each of the 6 candidates 2 simulated operas were run. The composer then listened to these and picked by ear his favourite of the 6 candidates. Figure 2 shows the finally selected tunes. The phrases are the "scaffold" from which a sung trade is constructed. They are very simple, and the construction is simple. The first pitch class is used for the buy/sell, the second for quantity, the third for the stock name, and the forth for the price.

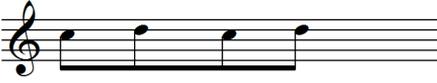
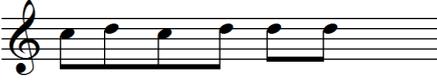
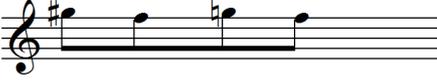
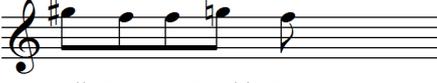
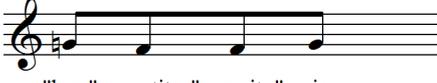
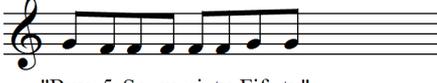
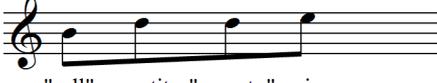
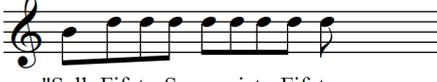
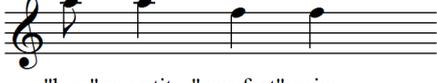
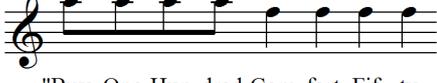
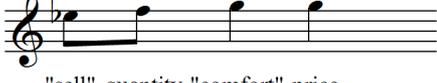
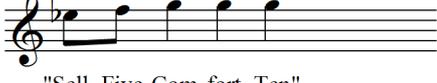

**Figure 2.** Examples of how trading phrases are used

The best way to explain the process is by examples, shown in Figure 2. It can be seen that the process involves repeating the "scaffold pitch" with the duration of the "scaffold duration". As can be seen, the words for trading a stock are said or sung in this order:

(1) "Buy" or "Sell"

(2) Quantity requested (e.g. "5", "11", "15", "27").

(3) Stock ("Wealth", "Comfort" or "Protection")

(4) Price request ("3.5", "7", "100")

So for example if you want to own 5 units of Wealth and are willing to pay at most 3.5 per unit, the order is "Buy 5 Wealth 3.5". If you want to get rid of 27 units of Protection and want to get at least 100 per unit then your order would be "Sell 27 Protection 100". Most of the time these words are sung. However there are times the conductor will signal that they can only be shouted (as in traditional open outcry markets). A trading phrase can be sung in any octave that is comfortable for the singer. They have to sing at the visual metronome time (controlled by the conductor), or at any multiple or divisor of that time (e.g. double speed, half speed etc).

As with real open outcry markets, to help with communicating orders to other trader-singers, hand signals are also used. The Buy / Sell signals are the same as those used in real markets, as shown in Figure 3. The hand signals for the three stocks are:

- Wealth: hold hands together (like rubbing hands "in glee").
- Protection: Wrap arms around body hugging yourself
- Comfort: Put palms together like praying and lean side of head on them – like symbol for falling asleep on your hands

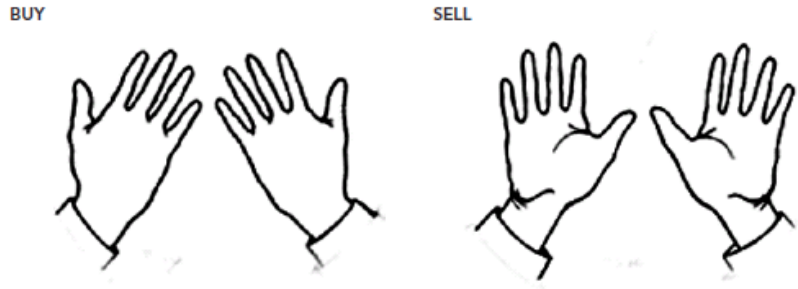

**Figure 3.** Buy / Sell hand Signals

Quantity and price are signaled in ways similar to those used in real open outcry markets, but in a simplified way. These were work-shopped during rehearsals.

The performance opens with a pre-scored cello overture by the composer, followed by the conductor ringing a real bell once to signify the market opening. The cello provides a tuning cue at market open (and at regular intervals throughout the performance). Each performer will have been assigned a portfolio with a certain quantity of the three "stocks". As mentioned, any profits they make on this portfolio, contribute to deciding the amount of money they are paid for their performance.

On 6 large screens around the performance venue, trades made can be seen as well as the content and value of portfolios. The prices of the three stocks are also visible, as well as graphs of their movements, and news items on screens. These news items can sometimes be clues to which way the market is going to move. Example news items include "Investors increase allocation to risky assets", "Investment sentiment balanced as GDP expectations unchanged", and "Global demand slumps, sharply reducing asset price expectations". These were generated automatically by the system, depending on which market state it was going to move into (boom, bust, or normal).

If another singer wants to take a trade being sung, they must sing the "opposite side". For example if the first singer wants to buy wealth, the other trader will sing the sell wealth tune. The 2 traders have to sing in exact time and phase with each other at least once. Once that is done the trade is agreed. Then they must try to attract the attention of the market administrators who will be visible on the floor. To do this, they point at each other and both must sing (in the same tune scaffold they traded with) "Re-con-cile Us" (four notes) repeatedly and in time until a market administrator can get to them. When the market administrator gets to a trader, the trader must detail their trade on a cardboard slip and pass it to the administrator. The administrator will also speak to their counterparty. Providing the two trades match the trade will be entered into the market, and displayed on the screens.

The main hand signals provided by the conductor are to start placing orders, and to switch in and out of shouting mode.

One element that became clear during development, workshopping and rehearsals was that the tempo of the performance needed to be kept sufficiently low to allow trader/singers to deal with the complexity of performing/trading. Furthermore it was seen that traders needed to advertise their offers multiple times to enable them to make a sale. Thus the final trading floor became more stylized – un-like the fast-paced and chaotic environment. This was not the originally envisioned atmosphere. However, it was considered more important to make the floor work as a trading arena, rather than to ensure it had the same level of sonic intensity as a real trading floor. As a way of partially counter-acting this effect, the conductor was able to signal to all singers to trade by shouting instead of singing. This freed them from the tempo restrictions, and for a segment of the performance there was a less stylized and fairly representative imitation of a real open outcry floor. The conductor also gradually increased the visual metronome speed by a few BPM as the performance progressed.

**Underlying Market**

It has been discussed how simulation is one method of bringing the "external world" into a reality opera in a dynamic way. There were three stocks and artificial cash in the underlying market simulator being run in real-time on a laptop. A standard random model of a stock market was used (Karatzas et al., 1998). Each stock had an "underlying" average return $R_i$. This represented how rapidly the stock price would go up or down – how profitable it could be. There will be an underlying market price $p_i$ for each stock which is displayed on a screen. This price is adjusted based on the current value of $R_i$. Each also had an underlying co-variance matrix $C_{ij}$. This modelled two elements: (1) how stable the stock price was while it moved up or down according to its returns; (2) how much the stock price was effected by the other two stocks in the market. Wealth had the highest return and volatility, Protection a little lower, and Comfort had the lowest returns but also the lowest risk.

As mentioned, the market model could be triggered into a Boom, Normal, or Bust behaviour by the conductor, and there was actually a probability of it switching between these three randomly. In each, the values of $R_i$ and $C_{ij}$ for each stock have different values – thus causing the stocks to behave in more positive, negative or neutral ways. In Boom all were likely to rise, and in Bust all were most likely to fall. The longer the market is in one type of state (boom, bust, normal) the greater the chance of it switching to another state. The possible interactions between Wealth, Comfort and Protection are fairly complex as the market moved between different states.

Each market state also had a set of random news stories that could occur in that state. The prices update every 15 seconds so the news story can be related directly to the next market state (which the system will 'know'/generate just after the previous price release, but before the prices update). Figure 4 shows a singer with one of the four sets of market screens in view, showing the stock price progressions and trader portfolios.

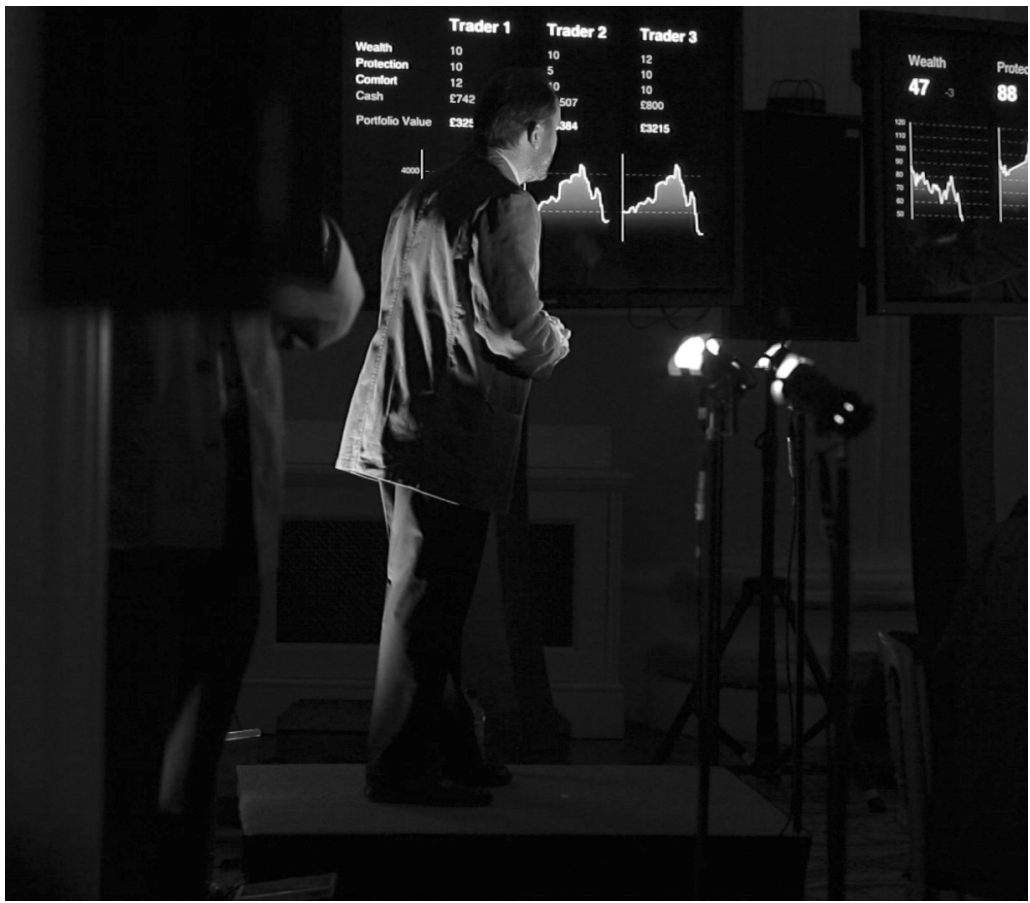

**Figure 4.** Picture of a Singer in front of Display Screens

There was a possibility that some singers would simply stop trading, to avoid the risk of losing money. So artificial hyperinflation was used, 0.2% every 15 seconds. If a singer horded their artificial cash, its value would gradually decrease. A final backup that was made available allowed a market administrator to give out a free unit of cash to all traders equally (£50). This was in case the traders performed so badly, that they all ran out of money and were in danger of stopping singing.

At the end of the 30 minute opera, the singer's portfolio is converted to cash based on the 'underlying' market prices, added to any cash they are holding, to give the final result. The highest scoring singer would then get a larger performance fee than the lowest scoring singer.

**Staging**

The opera was premiered at Mansion House – the residence of the Lord Mayor of the City of London (at the centre of the UK's financial business district). It was directed by Alessandro Talevi (of Sadlers Wells and Welsh National Opera). The normal recreation of an open outcry trading floor would have involved the singers standing close together in a small circle of 12. This would leave the audience to observe from the outside. Talevi believed this would leave the audience feeling excluded. He proposed the singer-traders are stand around the audience. This would enable the audience to experience the "vital dynamic of what it feels like to be within the ring of energy created by a circle of traders communicating". This approach looked more stylised than a standard open outcry floor, but had the advantage of immersing audience and helping them to view the "various threads of communication that make up the whole."

      Talevi also proposed that the staging should be as simple as possible. This was to allow the singers to focus on buying and selling stocks, and for the audience to be able to follow what was happening. Costume and lights should provide "a sense of theatricality in a delicate balancing act with the reality of a trading floor – because it is an actual (albeit artificial) trading floor, not a mere theatrical impression of one." Figure 5 shows one side of the trading floor at the premiere, with the conductor at the far left of frame.

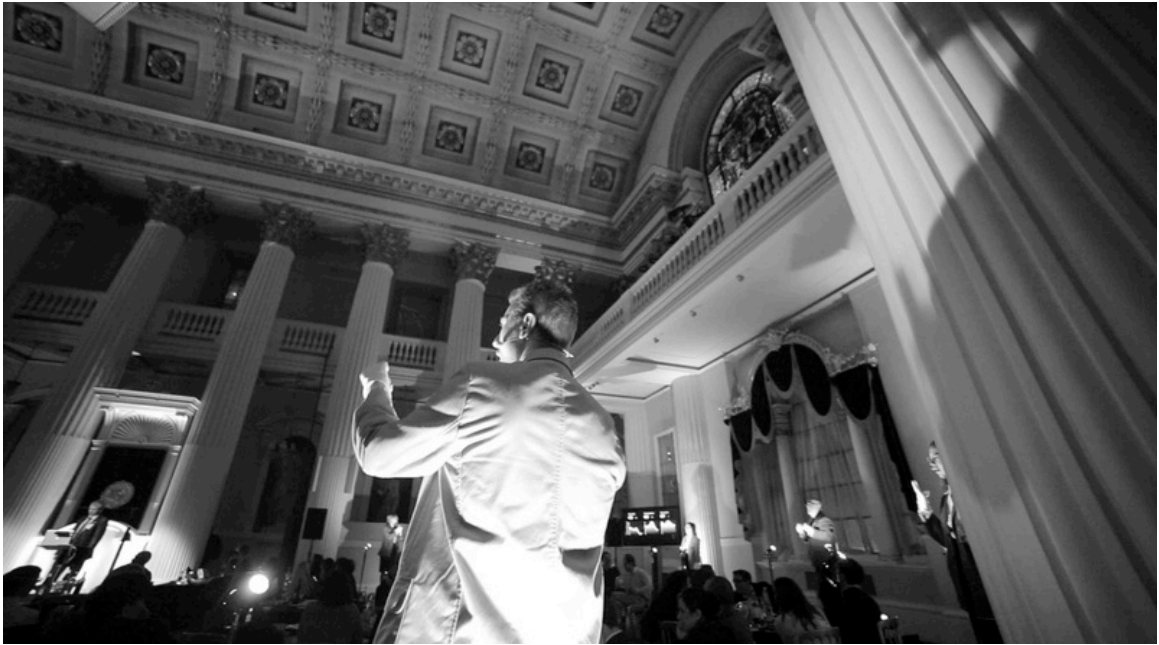

**Figure 5.** Some of the Trading Floor Staging for the Premiere

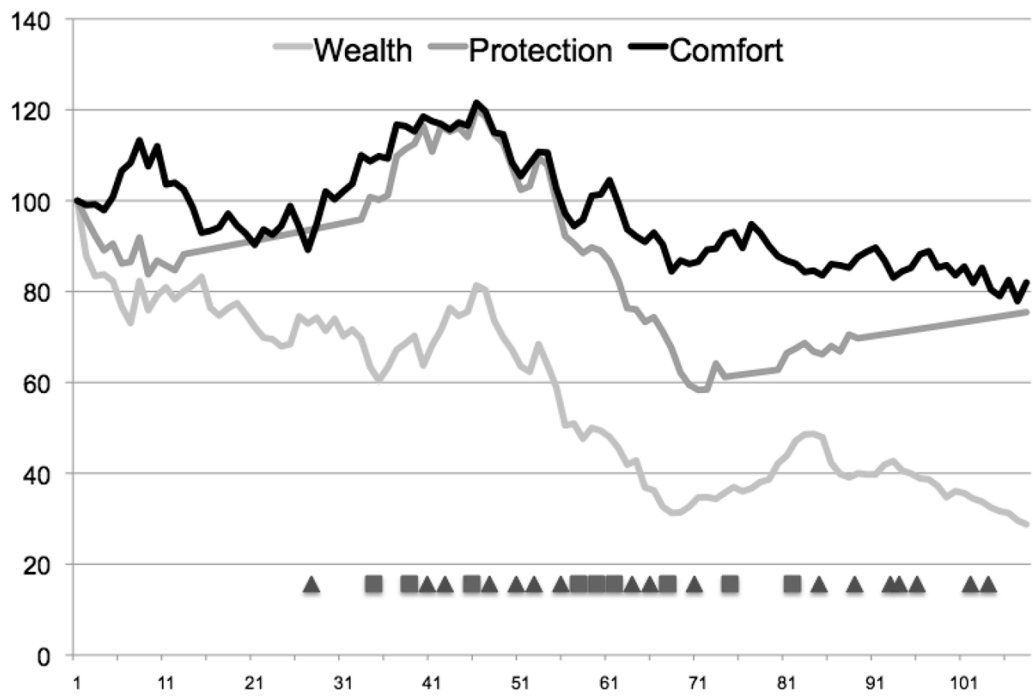

**Figure 6.** The Underlying Trading "Narrative" at the premiere.

**Interaction Results**

The actual trading and price "narrative" is shown in Figure 6 and some of the results are shown in Table 1. The market behaved fairly negatively. It was in Bust more than in any of the rehearsals. This included the enforced period of Bust early on as part of the secret "market score" controlled by the conductor. The final stock prices at the premiere were Protection 75, Wealth 29 and Comfort 82. Any singers who bet on a recovery and bought Wealth lost out. (Though, given the market model settings, in 70% of possible performances this would've been a good strategy.) The singer/trader whose portfolio had the highest value at the end was singer 1. They received 16.9% of the pot, as well as their performance fee. The singer whose portolio was the lowest value, (singer 12) was give 4.7% of the pot in addition to their fee.

| # | Number of Trades | Number buy | Number sell | Buy / sell proportions | Avg number of stocks traded | Number of trades initiated |
|---|---|---|---|---|---|---|
| 1 | 8 | 3 | 5 | 0.6 | 3.5 | 6 |
| 2 | 4 | 1 | 3 | 0.33 | 2.8 | 3 |
| 3 | 2 | 0 | 2 | 0 | 3.5 | 1 |
| 4 | 1 | 0 | 1 | 0 | 2 | 0 |
| 5 | 8 | 4 | 4 | 1 | 3.1 | 2 |
| 6 | 3 | 2 | 1 | 2 | 2 | 2 |
| 7 | 10 | 5 | 5 | 1 | 2.2 | 5 |
| 8 | 4 | 3 | 1 | 3 | 4 | 2 |
| 9 | 4 | 3 | 1 | 3 | 4.3 | 3 |
| 10 | 4 | 2 | 2 | 1 | 4.3 | 3 |
| 11 | 6 | 4 | 2 | 2 | 2.8 | 5 |
| 12 | 8 | 4 | 4 | 1 | 4 | 4 |
| **Avg** | **5.2** | **2.6** | **2.6** | **1.2** | **3.2** | **3** |

**Table 1.** Trading results from the premiere

Detailed feedback was provided by two-thirds of the performers (8 individuals). They were asked 5 specific questions, the results of which are shown in Figures 7 to 11. From Figure 7 it can be seen that ease of trading had an emotional impact on almost 90% of the sample, 38% agreeing the influence was strong.

For one of the sample, the ease of trading had less effect. This one trader repeats their response of "disagree" through all of Figures 7-11, but did not provide any details of why this would be so. They are perhaps an outlier, though they could also indicate that the methods used to rehearse the reality opera needed to be more personalised, requiring more rehearsals for individual interactions with the performers and director / composer.

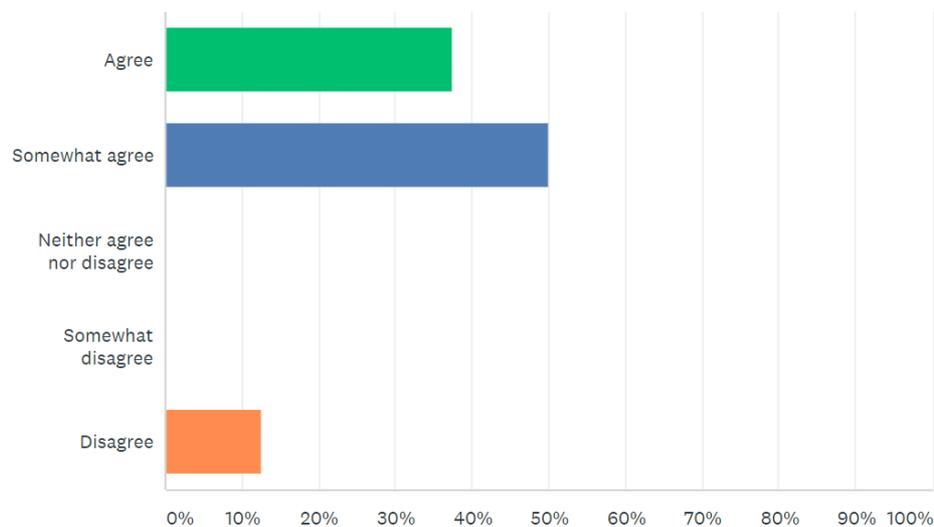

**Figure 7.** "My emotional state during the performance was strongly influenced by how easily I was able to trade."

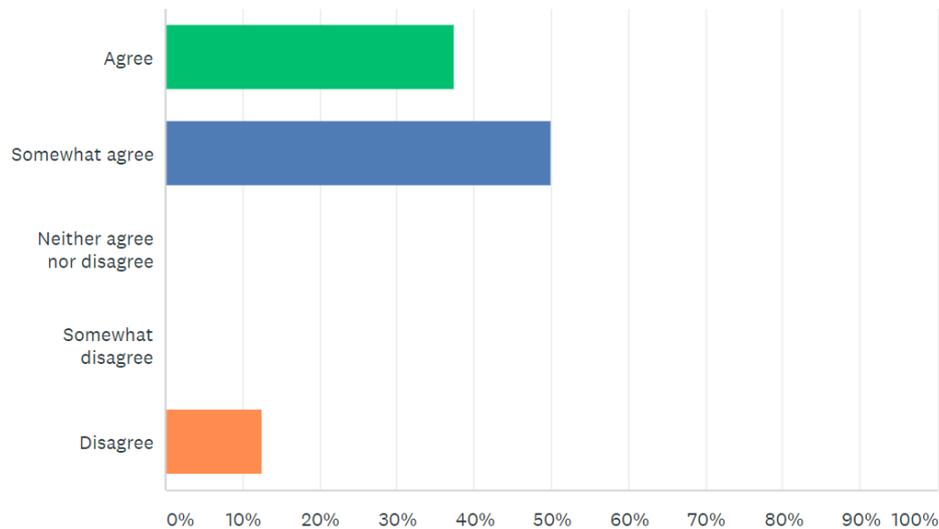

**Figure 8.** "My emotional state during the performance was strongly influenced by how well my portfolio was doing."

Figure 8 shows that most traders (almost 90%) agreed or somewhat agreed that their emotional state was strong influenced by how well their portfolio was doing, with 38% agreeing the influence was strong again. Figure 9 addresses part of the competitive element, not just asking if portfolio performance influenced real emotion, but if portfolio performance *relative to others* influenced real emotion. It is interesting to note that the proportions change here. Two of the singer traders move from agree or somewhat agree, to somewhat disagree. While that stills leaves 5 out of 8 saying their emotions were influenced or strongly influenced by others' portfolios, it suggests that traders were mostly taken up with keeping their own portfolios afloat and did not perhaps have the cognitive load (or sufficient rehearsal time) to take in clearly how others were doing. This is perhaps not surprising given there were 11 other portfolios to keep track of.

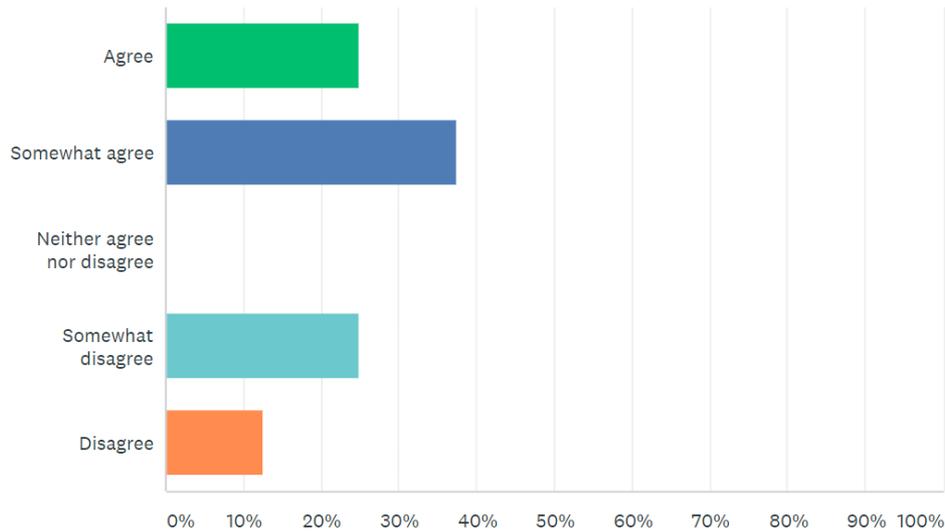

**Figure 9.** "My emotional state during the performance was strongly influenced by how well my portfolio was doing relative to others."

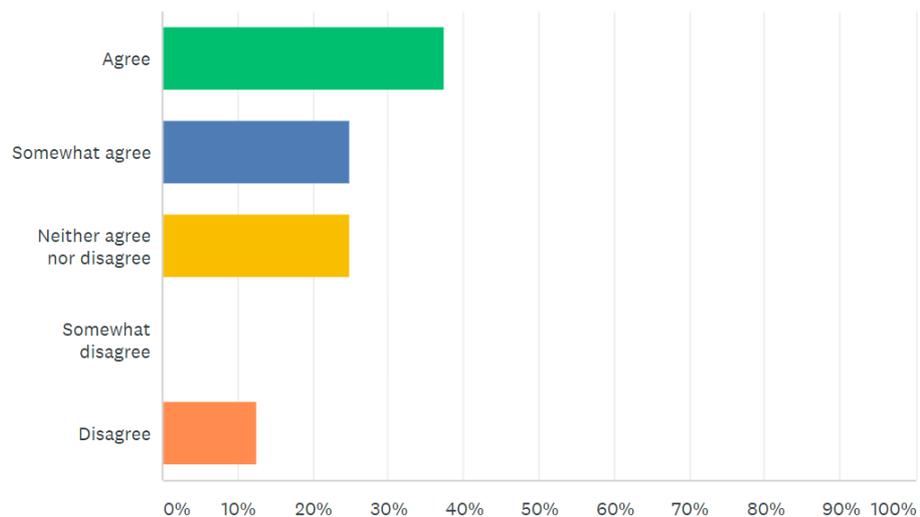

**Figure 10.** "My emotions during the performance were affected by the fact I knew there would be a loser and a winner, who would get lower and higher singers' fees."

In Figure 10 there is a similar shift of proportions regarding the potential prize money impact on real emotions. 5 out of 8 were influenced or strongly influenced emotionally by this. But 2 were now neutral. Finally Figure 11 asked the summary question to find out how much of the emotional experience was due to performance and music (as would be the case in traditional sung performance) and how much due to the dynamic self-direct narrative – the trading. 6 of the 8 agreed or somewhat agreed to the instrumental influence of the trading actions, 1 was neutral and the potential outlier disagreed again.

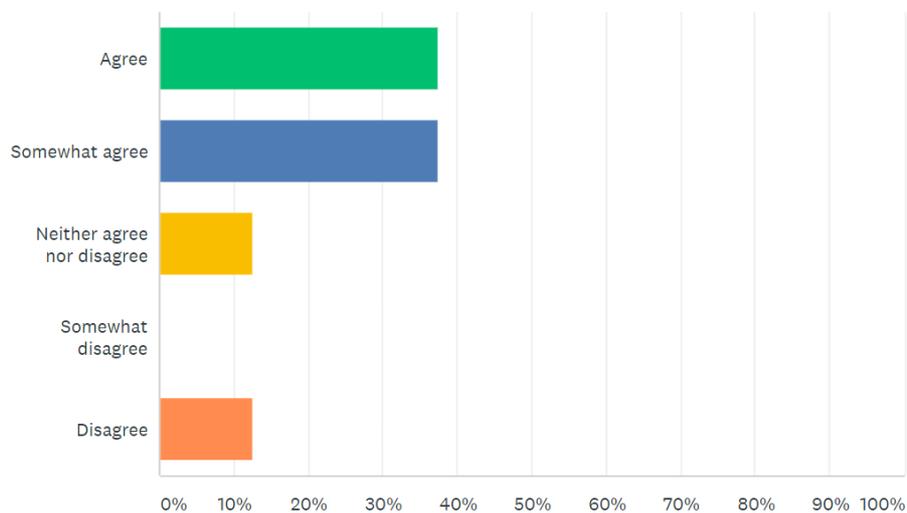

**Figure 11.** "Compared to other performances I have done, my emotions during Open Outcry were effected as much by what happened due to my trading actions, as they were by how well I performed or how the music itself made me feel."

When asked for freeform comments, some chose to respond. Here are comments from 3 of the singer/traders:

> "It was actually the spontaneous interaction from the audience, clapping and cheering, which made me suddenly realise, I was winning, and gave me the encouragement to keep going. The trading was fairly frustrating because you had to keep catching other traders' attention, do the trade by singing the correct motif, then write it down correctly, get the attention of the assistant to collect your slips of paper, all in the space of a few moments before the next trade. Cognitively very challenging but I relished it and the waves of sound from all the other traders, so intense. The audience reaction was the thing I didn't expect, and this was a wonderful part of the opera performance. It also so went against what we're 'allowed' to do in other opera, where the audience cannot clap and make their feelings known throughout the performance."

> "My feeling was that we were like 'day traders' - professional singers yes, but amateur traders dipping our toe in for the first time. Intellectually I would like to perform this again, this would somewhat mitigate the tension of going in blind, but I expect would make it more exciting in making profits or losses."

> "One of the audience, a trader, said that they recognised the look of panic when failing to make trades! It was a fascinating experience."

A less formal survey was performed on a significant proportion of the audience following the premiere. To summarise: they responded that they were very clear about what was happening (this would have been facilitated by an introductory talk and programme notes). In general they were somewhat surprised by the seeing the approach in action, and surprised by how much they enjoyed it. The musical experience was felt to be challenging, particularly to the audience members who were not regular attenders at new music events. However they still found the experience immersive and compelling.

The key element of reality opera described earlier was "emotions of the singers are real and emerge as a consequence of their reaction to a dynamic narrative." Although the above data is from only 8 of the 12 performers, and relates to a single performance, it supports that Open Outcry is a reality opera, and thus the plausibility of creating reality opera.

The data also supports that the performers were able to effectively trade using the system as devised. But it is clear that they found it cognitively challenging. For a future revision it would be necessary to have significantly more rehearsals. As far as audience understanding of what was happening, they were not surveyed. However the performer's observation above concerning applause for the audience near himself makes it clear that the system is capable of being understood by some of the audience.

**Conclusions**

This paper has proposed the concept of a reality opera. A number of solutions to crossing the barrier between opera and reality opera were proposed: (i) the restriction of the narrative to situations involving formalized human interactions, (ii) the use of methodologies from computer optimization to develop and experiment with the constraints, and (iii) the use of computer simulation to include elements of the outside world into the narrative and provide a testbed for the composer. An instantiation of the reality opera concept was produced and tested based on open outcry stock markets – a formalized human interaction. Genetic Algorithms and multi-agent modelling were used – referencing points (ii) and (iii) – in the composition of the reality opera. Although the simplification required because of the small number of rehearsals meant that GAs might not have been entirely necessary in this example. In reference to point (iii), an underlying dynamic market and news model was used in the example. After the performance, performers were surveyed and the resulting data – although based on 8 out of the 12 performers and from a single performance - supported that the performance, through the implementation of (i) to (iii), did indeed fulfill the requirements of a reality opera. It is distinct from previous work in non-deterministic sung musical ensemble performance, and also in comparison to previous musical and performance work using stock market dynamics. A video about the reality opera's premiere is available (Barclays, 2012), as is a recording of the full opera (Kirke, 2018).

The key element that came from the experience in relation to future reality operas was the need for more workshopping and rehearsals. The workshopping led to ways of using the cello to increase musical interest. Further workshopping would have been helpful in increasing the underlying musical meaningfulness of the performance. Listening to the recording of the performance shows its relative musical simplicity. However the audience were enthusiastic on the night – including a music critic for a national paper. This is explained by a combination of the concept with the music. To bring the complexity of the music up to the complexity of the concept would require significantly more workshopping and development. There is clearly much room in the area of reality opera for improvement.

Performers were working at their limit. Some perhaps beyond their limit. Doubling the number of rehearsals would have helped. Whether based in the world of markets, dating, sport or games, reality opera needs a longer rehearsal time compared to a traditional opera of the same length - partly because reality opera can take many paths. Like a quantum state, the performers and composer cannot know which path will be "collapsed to" on the night. Thus they must work through many possible paths, pointing to potential cost implications for reality opera as well.

Acknowledgements: Open Outcry was originated and composed by Alexis Kirke, and co-created by Alexis Kirke with Greg B. Davies (formerly head of behavioural finance at Barclays Bank). The underlying market model was coded by Joel Eaton. The performance was supported by Barclays Bank.

**References**

1. Barclays. 2012. "Open Outcry - Highlight video". Accessed 22 October 2018. http://www.youtube.com/watch?v=qx43RALtO8g
2. Bonardi, A., and Rousseaux, F. 2002. "Composing an Interactive Virtual Opera: The Virtualis Project." *Leonardo* 35(3), 315-318.
3. Chilver, P. 1967. *Improvised Drama*. London: Batsford Limited.
4. Ciardi, F. and At, F. 2013. "sMAX: A Multimodal Toolkit for Stock Market Data Sonification," In *Proceedings of the 10th International Conference on Auditory Display*, Sydney: ICAD.
5. Cohen, J. 1994. "Monitoring Background Activities." In *Auditory Display: Musification, Audification, and Auditory Interfaces,* 499. Santa Fe: Santa Fe Institute.
6. Çorlu, M., Maes, P.J., Muller, C., Kochman, K. and Leman, M. 2015." The impact of cognitive load on operatic singers' timing performance." *Frontiers in psychology* 6, 429.
7. Çorlu, M., Muller, C., Desmet, F. and Leman, M. 2015. "The consequences of additional cognitive load on performing musicians." *Psychology of Music* 43(4), 495-510.
8. Earley, M., 2014. "After the Twilight of the Gods: Opera Experiments, New Media and the Opera of the Future." In *Opera in the Media Age: Essays on Art, Technology and Popular Culture*, 229. Jefferson: McFarland & Company.
9. Cook, N. and Hayashi, T. 2008. "The Psychoacoustics of Harmony Perception." *American Scientist* 96(4), 311-319.
10. Dissanayake, E. 2008. "If music is the food of love, what about survival and reproductive success?" *Musicae Scientiae* 12(1), 169-195.
11. Earp, S. and Maney, D. 2012. "Birdsong: is it music to their ears?" *Frontiers in Evolutionary Neuroscience* 4, 14.
12. Econn, C. 2007. The Boomerang Kid, Los Angeles.
13. Boyle, R. 2011." With the Dow Piano, You Can Listen to the Tune of the Stock Market." *Popular Science.* https://www.popsci.com/technology/article/2011-

01/dow-piano-you-can-hear-strains-stock-market-2010, Bonnier Corporation USA.
14. Emerald Suspension. 2006. "Playing the Market." Emerald Suspension Label.
15. Goldberg, D. 1989. *Genetic Algorithms in Search, Optimization and Machine Learning*. Boston: Addison-Wesley Professional.
16. Haar, J. 1962. "On musical games in the 16th century." *Journal of the American Musicological Society* 15(1), 22-34.
17. Henson, K. 2016. *Technology and the Diva*. Cambridge: Cambridge University Press.
18. Jamini, D. 2005. *Harmony and Composition: Basics to Intermediate*. Bloomington: Trafford Publishing.
19. Juslin, P. and Sloboda, J. 2009. *Handbook of Music and Emotion: Theory, Research, Applications*. Oxford: Oxford University Press.
20. Karatzas, I. and Shreve, S. 1998. *Methods of mathematical finance*. New York: Springer.
21. Kirke, A. and Miranda, E. 2012. "Application of Pulsed Melodic Affective Processing to Stock Market Algorithmic Trading and Analysis," In *Proceedings of the 9th International Symposium on Computer Music Modeling and Retrieval*, London: Springer.
22. Kirke, A. and Miranda, E. 2015. "A multi-agent emotional society whose melodies represent its emergent social hierarchy and are generated by agent communications." *Journal of Artificial Societies and Social Simulation* 18(2) 16.
23. Kirke, A. 2018. "Open Outcry Full Audio." Accessed 22 October 2018. https://www.youtube.com/watch?v=euv1EfZmQfE
24. Kondek, C. 2005. "Dead Cat Bounce." Theater am Halleschen Ufer, Berlin.
25. Liuni, M. and Morelli, D. 2006. "Playing music: an installation based on Xenakis' musical games." In *Proceedings of the 2006 working conference on Advanced visual interfaces*, 322. Venezia:ACM.
26. Machover, T. 1996. "The Brain Opera and active music. Memesis: The Future of Evolution." In *Proceedings of the 1996 Ars Electronica*, 300. Linz: Springer.
27. Machover, T. 2011. "Future Opera for Robots and People Too." *OpenMind*, Accessed 24 October 2018, https://www.bbvaopenmind.com/en/articles/future-opera-for-robots-and-people-too/
28. Moon, B. and Moon, B. 2010. "Ear Trading."

29. Nyman, M. 1974. *Experimental Music: Cage and Beyond*. Cambridge: Cambridge University Press.
30. Rand, A. 1935. "Night of January 16th," Ambassador Theatre, New York.
31. Siegwart, H. and Scherer, K. 1995. "Acoustic concomitants of emotional expression in operatic singing: The case of lucia in Ardi gli incense." *Journal of Voice* 9(3) 249-260.
32. Studemann, F. 2010. "Music and Markets: Notes on a Crisis." *Financial Times*, December 21.
33. Ting, C.K., Wu, C.L., and Liu, C.H. 2017. "A novel automatic composition system using evolutionary algorithm and phrase imitation." *IEEE Systems Journal* 11(3), 1284-1295.
34. Turner, G. 2006. "The mass production of celebrity." *International Journal of Cultural Studies* 9(2) 153-165.
35. Waschka, R. 2007. "Composing with genetic algorithms: GenDash." In *Evolutionary Computer Music* 117. London: Springer.
36. Wooldridge, M. 2009. *An Introduction to Multi-agent Systems.* UK: Wiley and Sons.
37. Worrall, D. 2009. "The use of sonic articulation in identifying correlation in capital market trading data." In *Proceedings of the 15th International Conference on Auditory Display*. Copenhagen:ICAD.
38. Unander-Scharin, C. 2015. *Extending Opera - Artist-led Explorations in Operatic Practice through Interactivity and Electronics*, Doctoral Thesis, KTH Royal Institute of Technology, Stockholm.
39. Yun, B. 1989. *Cantonese Opera: Performance as Creative Process*. Cambridge: Cambridge University Press.